\documentstyle[11pt,epsf]{article}
\begin{document}
\begin{center}
{\Large\bf Probability of Incipient Spanning Clusters 
in Critical Square Bond Percolation}
\vspace{15mm}

Lev N. Shchur and Sergey S. Kosyakov

{\sl Landau Institute for Theoretical Physics, \\ 142432 Chernogolovka, Russia}

\vspace{20mm}

\begin{minipage}{100mm} \small \it
The probability of simultaneous occurence of at least $k$ spanning clusters
has been studied by Monte Carlo simulations on the 2D square lattice at
the bond percolation threshold $p_c=1/2$.
It is found that the probability of $k$ and more Incipient Spanning Clusters
(ISC) has the values $P(k>1) \approx 0.00658(3)$ and 
$P(k>2) \approx 0.00000148(21)$ provided that the limit of these
probabilities for infinite lattices exists. The probability $P(k>3)$
of more than three ISC could be estimated to be of the order of 
$10^{-11}$ and is beyond the possibility to compute a such value by
nowdays computers. So, it is impossible to check in simulations the
Aizenman law for the probabilities when $k>>1$.
We have detected a single sample with $4$ ISC in a total number of about
$10^{10}$ samples investigated. The probability of single event is
$1/10$ for that number of samples. 
\end{minipage}
\end{center}

\section{Introduction}
\noindent
It was a common belief until a very recent time 
that on 2D lattices at percolation threshold $p_c$
there exists exactly one percolation cluster \cite{Intro,Grimt}.
Indeed, it was rigorously proven by Newman and Shulman \cite{Inf01}
that the number of Infinite Clusters is either $0,\; 1$ or $\infty$
and later is was proven \cite{Uniq1,Uniq2} that the infinite cluster is
unique.

New insight developed recently by Aizenman,
who proved \cite{Aiz1} (see also, his talk at StatPhys 19
\cite{Aiz2}) that the number of Incipient Spanning Clusters (ISC) in 2D
critical percolation can be larger than one,
and that the probability of at least $k$ separate clusters is bounded

\begin{equation}
P_L(k) \left\{ \begin{array}{lr}
 \ge & A\; e^{-\alpha \  k^2} \\
 \le &     e^{-\alpha' \  k^2} 
\end{array} \right. 
\end{equation}
where $\alpha$ and $\alpha'$ are positive and $L$ is a linear lattice
size. Moreover, he conjectured the existence of the limit

\begin{equation}
\lim_{k\rightarrow\infty} \; \lim_{L\rightarrow\infty} \ \frac{1}{k^2} 
\log \; P_L(k) = \alpha_{asymp} \ .
\end{equation}

Indications of the existence of simultaneous clusters in two dimensional
critical percolation in the limit of infinite lattices  were found in
computer simulations of  Parongama Sen \cite{S1,S2} for site
percolation on square lattices~\footnote{Actually, P. Sen found more
than one  spanning cluster not only in dimension 2 but also in dimensions
3,4 and 5.} with helical boundary conditions
and in a strip geometry by Hu and Lin \cite{Strip}. The detailed history
of the recent development can be found in recent Stauffer mini-review
\cite{Mini}. 

Here we report on our attempt at a direct verification of Aizenman
rigorous result by computer simulations. 

We investigate by Monte-Carlo the number of spanning clusters 
in the critical bond percolation model on two dimensional square
lattices. We have determined the first limit in (2), i.e. the numerical
values of the probabilities 

\begin{equation}
P(k) =  \lim_{L\rightarrow\infty} \  P_L(k)  \ 
\end{equation}
for $k=1,2$ and $3$, as shown in Fig.~3.

Using calculated probabilities we can conclude 
that the probability $P(k=4)$ of four ISC is of the order
of $10^{-11}$ and is beyond the power of today's computers.
So, the numerical check of Aizenman result is a good task for the
computers of XXI century.

In the Section 2 we define precisely the model and the algorithm used in
simulations. The influence of subtle details of alghorithms on the values
of spanning probabilities  was
emphasized recently by Aharony and Stauffer \cite{Test}. 
In Section 3 we present the details of simulations.
The discussion of the results is in the last section.

\section{Model and Algorithm}

We use in simulations a rectangular square lattice with linear size $L$,
and exactly $L$ sites and $L$ bonds both horizontally and vertically.
We use free boundary conditions.
For clarity the example of lattice is shown in Fig. 1 together with the
dual lattice. The dual lattice has the same number of sites and bonds as
the original lattice (compare with \cite{LPPS} and \cite{Grimt}). 
This gives the possibility to keep in the finite lattices some properties
of infinite lattice. First of all, the number of bonds is exactly twice
the number of sites. Second, the self-duality is valid for any
finite lattice size. Third, the horizontal and vertical directions are
equivalent.

Each bond could be occupied with probability $q=1-p$ and closed with 
the probability $p$. Given the realization to each of $2\cdot L^2$ bonds to
be occupied formed the sample. Each sample could be
decomposed in clusters of connected occupied bonds. 
For that we use Hoshen-Kopelman \cite{Alg-HK}  algorithm, which is exact.
We are interested in the spanning properties of such a clusters. Namely, 
what is the probability that a cluster connects the opposite borders of
square and what is the number of disjoint spanning clusters?

An event $h$ that the cluster spans the lattice horizontally 
is an event that at least one of the left sites
and at least one of the right bonds are in the same cluster. 
The probability of such event is just the Langlands et all \cite{LPPS}
horizontal crossing probability $\pi_h$.

For our purposes we need more detailed events. Namely, by $h_1$
we will denote the event that there are exactly one cluster connecting
left sites with the right occupied bonds. In the same manner by $h_k$
we will denote the event that there are exactly $k$ disjoint clusters 
connecting left and right borders of our lattice. 
Then, the horizontal crossing probility $\pi_h$ is given by
$\sum_{i=1}^\infty \pi_{h_i}$, where we denote by $\pi_{h_k}$ the
probability of event $h_k$.

In a full analogy, by $\pi_{v_k}$ we will denote the crossing probabilities
from top to bottom, i.e. vertically. Obviously, for our choice of
the lattice, $\pi_{h_k}=\pi_{v_k}$ for any lattice size. This could be
used  as a check of the calculated statistical errors.

Knowing the origin of the bad properties of both main fast methods for
random number generation \cite{Mars,SHB} we check the results using the
same linear congruential method as Langlands, et all use in \cite{LPPS} 
($x_{i+1}=(a\; x_i +c) \ {\rm mod}\; m$, with $a=142412240584757$, 
$c=11$, $m=2^{48}$) and the shift register
($x_n=x_{n-157} \oplus x_{n-314} \oplus x_{n-471} \oplus x_{n-9689}$) 
\cite{At}  (the one used by Ziff in \cite{Ziff}) 
and found that the results coincide, which suggests the absence 
of systematic errors.

It should be noted that the variance of probabilities is independent from
tha lattice size because probabilities are calculated as the expectation
values of corresponding indicator functions. 
So, we should keep the number of samples independent of the lattice
size and, therefore,  the only parameter which controls the
statistical errors is the number of samples $M$. Probabilities were
calculated averaging $M=10^8$ samples and the error bars were defined
from $100$ bins, each bin being the average over $10^6$ samples.

Throughout the next section we deal with the critical bond percolation
on square lattice ($p=p_c=1/2$) with free boundaries.

The simulations was done under the Topos environment \cite{Top1}
working on a number of Digital Alpha workstations and servers in Landau
Institute and Chernogolovka Science Park.

\section{Numerical results}

We calculate the probabilities $\pi_{h_k}$ and $\pi_{v_k}$
for square lattices with linear sizes $L=8,12,16,20,30,32,64$ looking 
for all events up to $8$.\footnote{This does not imply that we seriously 
expect any events with $k\ge 4$. The upper limit of eight is simply
related to the numerical algorithm.}
Within the statistical errors the vertical and horizontal crossing
probabilities coincinde giving additional confidence in the
quality of data (Table I).

  On the Fig.1 we plot the probability for simultaneous occurence of more
than one spanning cluster $P_L(k>1) = \sum_{k>1} \pi_{h_k}$
versus the inverse system volume $1/L^2$.  A linear fit gives us the
limiting value of $P(k>1)=6.58\cdot 10^{-3}$ with error $\approx 3\cdot 10^{-5}$.

 Fig.2 shows the dependence of the probability of more than two clusters 
versus $1/L^2$. The best linear fit gives the limiting value of
$P(k>2)=1.48\cdot 10^{-6}$ with uncertainty $2.1\cdot 10^{-7}$.

 Actually, we observed in computations mostly the events of up to
three simultaneous spanning clusters.  We simulated at total 
about $10^{10}$ samples of different sizes $8 \le L \le 64$
and only one sample with $4$ spanning clusters was detected. This event
clearly not contradicts with our estimate for $P(4)\approx 10^{-11}$, 
given below. Single event is probable as one part in ten.

 We could fit exponent $\alpha$ in (2) taking 
$\alpha_{k}=-1/k^2 \ \log \; P(n\ge k)$
and obtaining $\alpha_1=0.693$, $\alpha_2=1.256$ and $\alpha_3=1.498$.

 The logarithm of $P(k)$ is plotted on Fig.3 versus 
$k^2$ together with the best linear fit, which gives $\alpha \propto
1.61(7)$ and $A\propto 3\pm1.5$. For comparison, a linear fit for
the logarithm of $P(k)$ versus $k$ gives the value of exponent $2.8(5)$ 
with the too large uncertainty.

This is as full analysis as we could do because even with
the using the of $\alpha=1.5$ the probability
of four ISC would be $3.8\cdot 10^{-11}$ and
of five  ISC of $5.5 \cdot 10^{-17}$. Even the first value is
impossible to check with the computers one have today. 
To estimate accurately the probability of an event 
expected to be of the order $10^{-11}$ one needs to generate at least
$10^{13}$ samples. Suppose, we have a computer with the CPU cycle of
$\Delta$ ns and an algorithm which needs $m$ CPU cycles to process one
lattice site.
Thus, it is $m\cdot L^2$ cycles per sample and $10^{13} \cdot m\cdot L^2$ cycles
altogether. The time needed will be
$10^{13} \cdot m\cdot L^2\cdot \Delta$. If we are very optimistic, we could
think that $m=10$ and $\Delta=1$ ns, and taking a moderate lattice
size $L=32$ one gets about $1157$ ... days! 

In addition, we compute the distribution of cluster sizes. In Fig.4 we
plot the mean value $<s>$ per lattice site of ISC of
corresponding type. It is clear that the all clusters is of the same
type, i.e. described by the same exponent, which is 
$\beta/\nu=5/48=0.10416...$ as it should be [1] and as stressed in \cite{S2}. 
A linear fit gives the actual values 
$0.1056(14)$, $0.0999(20)$ and $0.1046(26)$, which
are not far from the exact value. This is the additional
argument that we are in force to summarize the probabilities $\pi_{h_k}$.

\section{Discussion}

We confirm more accurately the simulation result of \cite{S1,S2} 
\footnote{Actually, we got most of the results before getting
the last preprint}
for 2D percolation: the probability of two disjoint Incipient Spanning Clusters
has a small but finite value. The difference in the values for
probabilities is due to the fact that  in \cite{S1,S2} helical boundary
conditions are used. Whereas we used free boundary conditions.

The crossing probability with periodic boundary conditions (PBC) is known
to be larger and the value $0.63665(8)$ is computed in \cite{AH-PRE}.
We estimate in computer simulations the probability of
disjoint ISC clusters to be $P(k>1) \approx 2.0(4) \ 10^{-3}$ and
$P(k>2) \approx 1.4(5)\ 10^{-7}$ which are even smaller than for the
case of free boundary conditions (FBC) considered by us.
The finite size scaling of PBC is more complicated \cite{AH-PRE}
and fit of the results for only four lattice sizes (8,16,32,64) we
simulated gives less accurate limiting values in comparison with the
one for FBC. Surprisingly, the linear fit of logarithm of $P(k)$
versus $k^2$ (shown in Fig. 4 by solid line) gives better accuracy 
then for the case of FBC: $\alpha_{PBC} \approx 1.915(1)$ and 
$A\approx 4.26(3)$.

It is necessary to note, that we assume the existence of the limiting
probabilities (2) throughout this work and that the probabilities
$\pi_{h_k}$ with $k>1$ reach their maximum in critical region.
Indeed, our preliminary simulations \cite{SKV} shows that the maximum of 
probabilities $\pi_{h_1}$ and $\pi_{h_2}$ occurs at some $p_{max}$,
which varies as $(p_{max}-p_c) \propto  L^{-1/\nu}$ ($\nu=4/3$ is the
correlation length exponent) and that the limiting
probabilities computed here seems to be correct.

Occasionally, from a total number of $10^{10}$ samples investigated by
us (not all the data presented here), we detect the one event of sample
with just four Incipient Spanning Clusters, which is as probable
event as $1/10$. 

\section{Acknowledgments}

Authors are grateful to D. Stauffer for valuable spanning 
discussions of the problem and for sending us preprints  \cite{S2,Test,Mini}
prior publication. 
The carping discussions with Oleg Vasilyev are very appreciated. 
Many thanks to P. Butera for discussion and careful reading
of the draft.
One of us (LNS) is thankful for hospitality to the Theoretical Physics
Group of Milano University, where he stays as CARIPLO Fellow and
where this work was finished,  and especially to P.Butera,
M. Comi, G. Marchesini and G. Salam for help in various occasions.
This work is supported in part by grants RFBR 96-02-18168, 
NWO 07-210 and INTAS 93-211. LNS is supported by the Cariplo Foundation
for Scientific Research.

\newpage

\begin{table}
\caption{Probabilities of more than $k$ incipient spanning clusters
on critical bond square lattices with linear size $L$ and free
boundaries. Note, that
the values of $P(k>1)$ is multiplyed by a factor $10^3$ and those
of $P(k>2)$ by a factor $10^6$. For each lattice size $L$ the first
row is the probability of horizontal crossing and the second one is the
probability of vertical crossing.}
\vspace{5mm}

\begin{tabular}{r|r|r|r} \hline
L   & $k>0$      & $k>1$ $10^3$ & $k>2$ $10^6$ \\ \hline \hline
8   & 0.50005(5) & 7.657(8)     & 3.40(15) \\ \hline
    & 0.50003(4) & 7.660(8)	& 3.98(21) \\ \hline\hline
12  & 0.50002(5) & 7.084(9)	& 2.57(14) \\ \hline
    & 0.49995(5) & 7.070(8)	& 2.10(13) \\ \hline\hline
16  & 0.50003(7) & 6.855(9)	& 1.97(17) \\ \hline
    & 0.50002(6) & 6.843(8)	& 1.79(19) \\ \hline\hline
20  & 0.49990(6) & 6.742(8)	& 1.95(14) \\ \hline
    & 0.50008(5) & 6.745(9)	& 1.72913) \\ \hline\hline
30  & 0.49999(4) & 6.650(8)	& 1.52(14) \\ \hline
    & 0.49996(5) & 6.653(7)	& 1.52(12) \\ \hline\hline
32  & 0.49999(5) & 6.648(8)	& 1.73(12) \\ \hline
    & 0.50008(7) & 6.642(8)	& 1.56(11) \\ \hline\hline
64  & 0.49992(9) & 6.597(9)	& 1.33(13) \\ \hline
    & 0.49999(6) & 6.602(8)	& 1.51(14) \\ \hline\hline
$\infty$ & 0.50002(2) & 6.58(3)   & 1.48(21) \\       
\end{tabular}
\end{table}

 
\begin{figure}
\centering
\epsfxsize=110mm
\epsffile{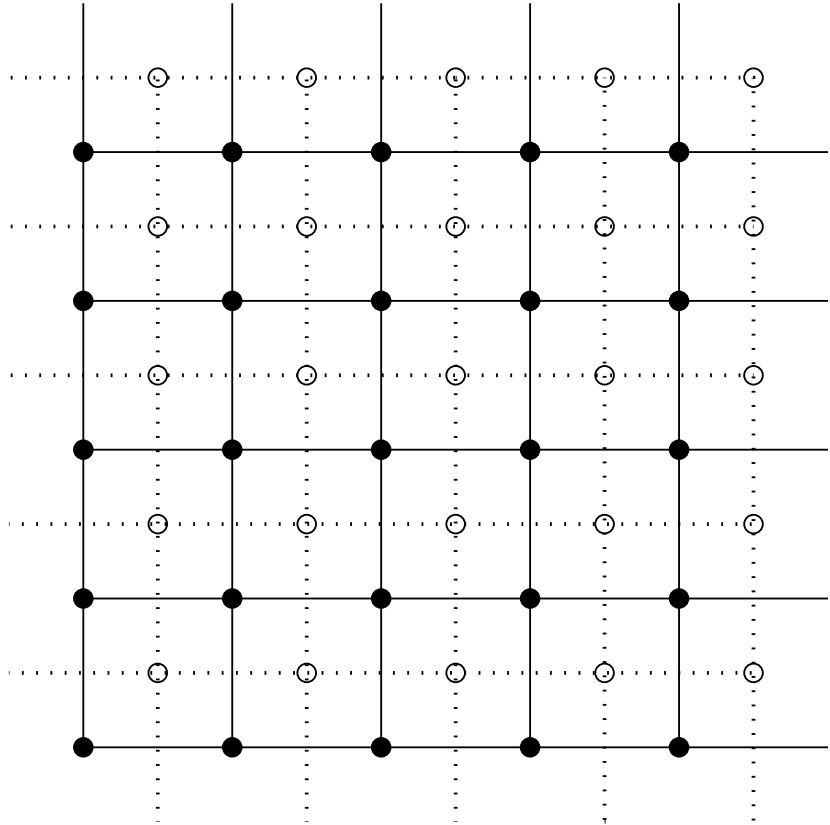}
\vskip 5mm
\caption{Example of simulated lattice with linear size $L=5$ (solid lines) 
and its dual (dotted lines). Note, that the number of sites and bonds in
both directions and for both lattices is just equal to $L$.
}
\label{Conv}
\end{figure}

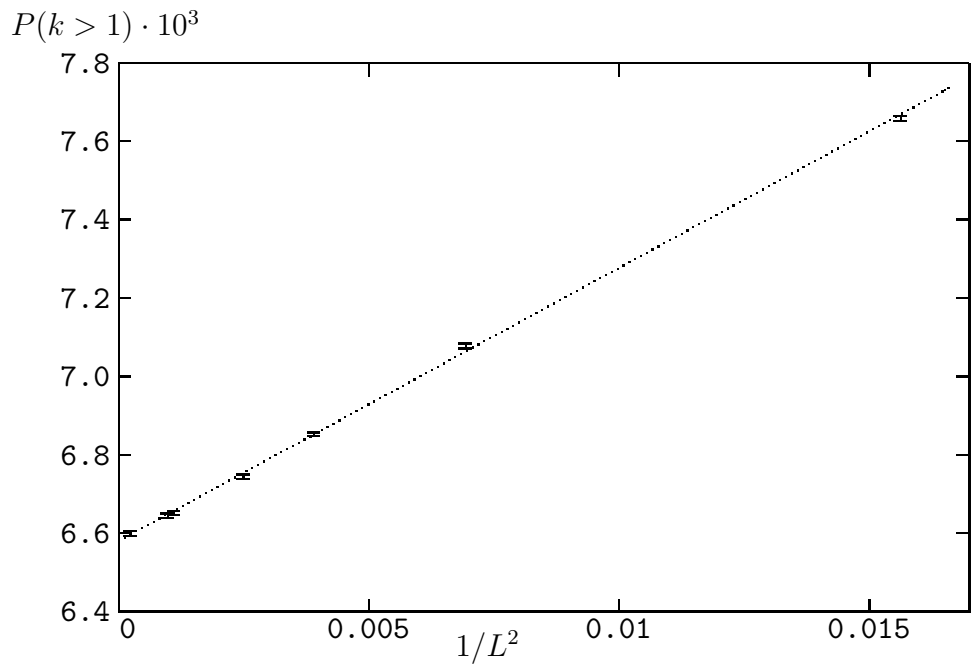
\begin{figure}
\centering
\setlength{\unitlength}{0.240900pt}
\ifx\plotpoint\undefined\newsavebox{\plotpoint}\fi
\begin{picture}(1500,900)(0,0)
\font\gnuplot=cmtt10 at 12pt
\gnuplot
\sbox{\plotpoint}{\rule[-0.200pt]{0.400pt}{0.400pt}}%
\put(120.0,31.0){\rule[-0.200pt]{0.400pt}{207.656pt}}
\put(120.0,31.0){\rule[-0.200pt]{4.818pt}{0.400pt}}
\put(108,31){\makebox(0,0)[r]{6.4}}
\put(1436.0,31.0){\rule[-0.200pt]{4.818pt}{0.400pt}}
\put(120.0,154.0){\rule[-0.200pt]{4.818pt}{0.400pt}}
\put(108,154){\makebox(0,0)[r]{6.6}}
\put(1436.0,154.0){\rule[-0.200pt]{4.818pt}{0.400pt}}
\put(120.0,277.0){\rule[-0.200pt]{4.818pt}{0.400pt}}
\put(108,277){\makebox(0,0)[r]{6.8}}
\put(1436.0,277.0){\rule[-0.200pt]{4.818pt}{0.400pt}}
\put(120.0,400.0){\rule[-0.200pt]{4.818pt}{0.400pt}}
\put(108,400){\makebox(0,0)[r]{7.0}}
\put(1436.0,400.0){\rule[-0.200pt]{4.818pt}{0.400pt}}
\put(120.0,524.0){\rule[-0.200pt]{4.818pt}{0.400pt}}
\put(108,524){\makebox(0,0)[r]{7.2}}
\put(1436.0,524.0){\rule[-0.200pt]{4.818pt}{0.400pt}}
\put(120.0,647.0){\rule[-0.200pt]{4.818pt}{0.400pt}}
\put(108,647){\makebox(0,0)[r]{7.4}}
\put(1436.0,647.0){\rule[-0.200pt]{4.818pt}{0.400pt}}
\put(120.0,770.0){\rule[-0.200pt]{4.818pt}{0.400pt}}
\put(108,770){\makebox(0,0)[r]{7.6}}
\put(1436.0,770.0){\rule[-0.200pt]{4.818pt}{0.400pt}}
\put(120.0,893.0){\rule[-0.200pt]{4.818pt}{0.400pt}}
\put(108,893){\makebox(0,0)[r]{7.8}}
\put(1436.0,893.0){\rule[-0.200pt]{4.818pt}{0.400pt}}
\put(120.0,31.0){\rule[-0.200pt]{0.400pt}{4.818pt}}
\put(135,5){\makebox(0,0){0}}
\put(120.0,873.0){\rule[-0.200pt]{0.400pt}{4.818pt}}
\put(513.0,31.0){\rule[-0.200pt]{0.400pt}{4.818pt}}
\put(513,5){\makebox(0,0){0.005}}
\put(513.0,873.0){\rule[-0.200pt]{0.400pt}{4.818pt}}
\put(906.0,31.0){\rule[-0.200pt]{0.400pt}{4.818pt}}
\put(906,5){\makebox(0,0){0.01}}
\put(906.0,873.0){\rule[-0.200pt]{0.400pt}{4.818pt}}
\put(1299.0,31.0){\rule[-0.200pt]{0.400pt}{4.818pt}}
\put(1299,5){\makebox(0,0){0.015}}
\put(1299.0,873.0){\rule[-0.200pt]{0.400pt}{4.818pt}}
\put(120.0,31.0){\rule[-0.200pt]{321.842pt}{0.400pt}}
\put(1456.0,31.0){\rule[-0.200pt]{0.400pt}{207.656pt}}
\put(120.0,893.0){\rule[-0.200pt]{321.842pt}{0.400pt}}
{\large
\put(100,950){\makebox(0,0){$P(k>1) \cdot 10^3$}}
\put(700,-25){\makebox(0,0){$1/L^2$}}
}
\put(120.0,31.0){\rule[-0.200pt]{0.400pt}{207.656pt}}
\put(1348.0,802.0){\rule[-0.200pt]{0.400pt}{1.686pt}}
\put(1338.0,802.0){\rule[-0.200pt]{4.818pt}{0.400pt}}
\put(1338.0,809.0){\rule[-0.200pt]{4.818pt}{0.400pt}}
\put(665.0,444.0){\rule[-0.200pt]{0.400pt}{1.927pt}}
\put(655.0,444.0){\rule[-0.200pt]{4.818pt}{0.400pt}}
\put(655.0,452.0){\rule[-0.200pt]{4.818pt}{0.400pt}}
\put(427.0,307.0){\rule[-0.200pt]{0.400pt}{1.204pt}}
\put(417.0,307.0){\rule[-0.200pt]{4.818pt}{0.400pt}}
\put(417.0,312.0){\rule[-0.200pt]{4.818pt}{0.400pt}}
\put(316.0,239.0){\rule[-0.200pt]{0.400pt}{1.686pt}}
\put(306.0,239.0){\rule[-0.200pt]{4.818pt}{0.400pt}}
\put(306.0,246.0){\rule[-0.200pt]{4.818pt}{0.400pt}}
\put(207.0,182.0){\rule[-0.200pt]{0.400pt}{1.686pt}}
\put(197.0,182.0){\rule[-0.200pt]{4.818pt}{0.400pt}}
\put(197.0,189.0){\rule[-0.200pt]{4.818pt}{0.400pt}}
\put(197.0,178.0){\rule[-0.200pt]{0.400pt}{1.686pt}}
\put(187.0,178.0){\rule[-0.200pt]{4.818pt}{0.400pt}}
\put(187.0,185.0){\rule[-0.200pt]{4.818pt}{0.400pt}}
\put(139.0,150.0){\rule[-0.200pt]{0.400pt}{1.686pt}}
\put(129.0,150.0){\rule[-0.200pt]{4.818pt}{0.400pt}}
\put(129.0,157.0){\rule[-0.200pt]{4.818pt}{0.400pt}}
\put(120,142){\usebox{\plotpoint}}
\multiput(120,142)(9.109,4.9725){144}{\usebox{\plotpoint}}
\put(1417,850){\usebox{\plotpoint}}
\end{picture}
\vskip 5mm
\caption{Probability of more than one Incipient Spanning Cluster
multiplied by one thousand for 2D bond percolation model as a
function of $1/L^2$. The linear lattice sizes
$L$ are 8,12,16,20,30,32,64.
The probability approaches the value of
0.00658(53) in the limit of infinite $L$. Error bars are computed 
over 100 bins of $10^6$ samples each.}
\label{Prob2}
\end{figure}

\begin{figure}
\centering
\setlength{\unitlength}{0.240900pt}
\ifx\plotpoint\undefined\newsavebox{\plotpoint}\fi
\sbox{\plotpoint}{\rule[-0.200pt]{0.400pt}{0.400pt}}%
\begin{picture}(1500,900)(0,0)
\font\gnuplot=cmtt10 at 12pt
\gnuplot
\sbox{\plotpoint}{\rule[-0.200pt]{0.400pt}{0.400pt}}%
\put(120.0,31.0){\rule[-0.200pt]{0.400pt}{207.656pt}}
\put(120.0,31.0){\rule[-0.200pt]{4.818pt}{0.400pt}}
\put(108,31){\makebox(0,0)[r]{1.0}}
\put(1436.0,31.0){\rule[-0.200pt]{4.818pt}{0.400pt}}
\put(120.0,175.0){\rule[-0.200pt]{4.818pt}{0.400pt}}
\put(108,175){\makebox(0,0)[r]{1.5}}
\put(1436.0,175.0){\rule[-0.200pt]{4.818pt}{0.400pt}}
\put(120.0,318.0){\rule[-0.200pt]{4.818pt}{0.400pt}}
\put(108,318){\makebox(0,0)[r]{2.0}}
\put(1436.0,318.0){\rule[-0.200pt]{4.818pt}{0.400pt}}
\put(120.0,462.0){\rule[-0.200pt]{4.818pt}{0.400pt}}
\put(108,462){\makebox(0,0)[r]{2.5}}
\put(1436.0,462.0){\rule[-0.200pt]{4.818pt}{0.400pt}}
\put(120.0,606.0){\rule[-0.200pt]{4.818pt}{0.400pt}}
\put(108,606){\makebox(0,0)[r]{3.0}}
\put(1436.0,606.0){\rule[-0.200pt]{4.818pt}{0.400pt}}
\put(120.0,749.0){\rule[-0.200pt]{4.818pt}{0.400pt}}
\put(108,749){\makebox(0,0)[r]{3.5}}
\put(1436.0,749.0){\rule[-0.200pt]{4.818pt}{0.400pt}}
\put(120.0,893.0){\rule[-0.200pt]{4.818pt}{0.400pt}}
\put(108,893){\makebox(0,0)[r]{4.0}}
\put(1436.0,893.0){\rule[-0.200pt]{4.818pt}{0.400pt}}
\put(120.0,31.0){\rule[-0.200pt]{0.400pt}{4.818pt}}
\put(135,5){\makebox(0,0){0}}
\put(120.0,873.0){\rule[-0.200pt]{0.400pt}{4.818pt}}
\put(513.0,31.0){\rule[-0.200pt]{0.400pt}{4.818pt}}
\put(513,5){\makebox(0,0){0.005}}
\put(513.0,873.0){\rule[-0.200pt]{0.400pt}{4.818pt}}
\put(906.0,31.0){\rule[-0.200pt]{0.400pt}{4.818pt}}
\put(906,5){\makebox(0,0){0.01}}
\put(906.0,873.0){\rule[-0.200pt]{0.400pt}{4.818pt}}
\put(1299.0,31.0){\rule[-0.200pt]{0.400pt}{4.818pt}}
\put(1299,5){\makebox(0,0){0.015}}
\put(1299.0,873.0){\rule[-0.200pt]{0.400pt}{4.818pt}}
\put(120.0,31.0){\rule[-0.200pt]{321.842pt}{0.400pt}}
\put(1456.0,31.0){\rule[-0.200pt]{0.400pt}{207.656pt}}
\put(120.0,893.0){\rule[-0.200pt]{321.842pt}{0.400pt}}
{\large
\put(100,950){\makebox(0,0){$P(k>2)\cdot10^6$}}
\put(700,-25){\makebox(0,0){$1/L^2$}}
}
\put(120.0,31.0){\rule[-0.200pt]{0.400pt}{207.656pt}}
\put(1348,804){\raisebox{-.8pt}{\makebox(0,0){$\Diamond$}}}
\put(665,415){\raisebox{-.8pt}{\makebox(0,0){$\Diamond$}}}
\put(427,284){\raisebox{-.8pt}{\makebox(0,0){$\Diamond$}}}
\put(316,271){\raisebox{-.8pt}{\makebox(0,0){$\Diamond$}}}
\put(207,180){\raisebox{-.8pt}{\makebox(0,0){$\Diamond$}}}
\put(197,216){\raisebox{-.8pt}{\makebox(0,0){$\Diamond$}}}
\put(139,152){\raisebox{-.8pt}{\makebox(0,0){$\Diamond$}}}
\put(1348.0,767.0){\rule[-0.200pt]{0.400pt}{17.827pt}}
\put(1338.0,767.0){\rule[-0.200pt]{4.818pt}{0.400pt}}
\put(1338.0,841.0){\rule[-0.200pt]{4.818pt}{0.400pt}}
\put(665.0,387.0){\rule[-0.200pt]{0.400pt}{13.249pt}}
\put(655.0,387.0){\rule[-0.200pt]{4.818pt}{0.400pt}}
\put(655.0,442.0){\rule[-0.200pt]{4.818pt}{0.400pt}}
\put(427.0,252.0){\rule[-0.200pt]{0.400pt}{15.177pt}}
\put(417.0,252.0){\rule[-0.200pt]{4.818pt}{0.400pt}}
\put(417.0,315.0){\rule[-0.200pt]{4.818pt}{0.400pt}}
\put(316.0,243.0){\rule[-0.200pt]{0.400pt}{13.249pt}}
\put(306.0,243.0){\rule[-0.200pt]{4.818pt}{0.400pt}}
\put(306.0,298.0){\rule[-0.200pt]{4.818pt}{0.400pt}}
\put(207.0,154.0){\rule[-0.200pt]{0.400pt}{12.768pt}}
\put(197.0,154.0){\rule[-0.200pt]{4.818pt}{0.400pt}}
\put(197.0,207.0){\rule[-0.200pt]{4.818pt}{0.400pt}}
\put(197.0,192.0){\rule[-0.200pt]{0.400pt}{11.804pt}}
\put(187.0,192.0){\rule[-0.200pt]{4.818pt}{0.400pt}}
\put(187.0,241.0){\rule[-0.200pt]{4.818pt}{0.400pt}}
\put(139.0,124.0){\rule[-0.200pt]{0.400pt}{13.249pt}}
\put(129.0,124.0){\rule[-0.200pt]{4.818pt}{0.400pt}}
\put(129.0,179.0){\rule[-0.200pt]{4.818pt}{0.400pt}}
\put(120,147){\usebox{\plotpoint}}
\multiput(120,142)(9.109,4.9725){144}{\usebox{\plotpoint}}
\put(1417,810){\usebox{\plotpoint}}
\end{picture}
\vskip 5mm
\caption{Probability of more than two Incipient Spanning Cluster
multiplied by one million for 2D bond percolation model as a function 
of $1/L^2$. The linear lattice sizes
$L$ are 8,12,16,20,30,32,64.
The probability approaches the value of
0.00000148(21) in the limit of infinite $L$. Error bars are computed
over 100 bins of $10^6$ samples each.}
\label{Prob3}
\end{figure}
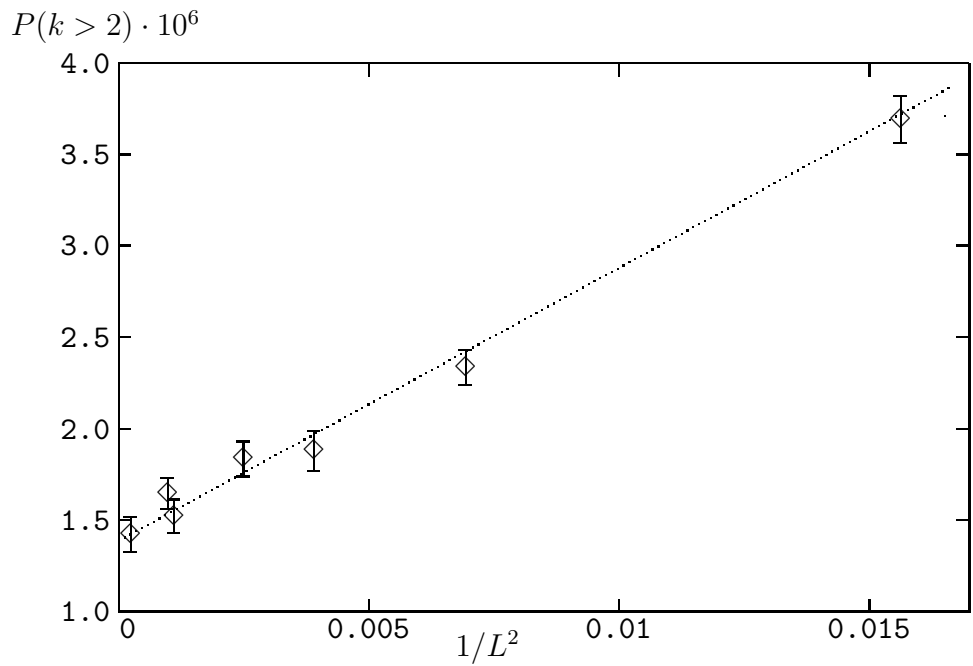

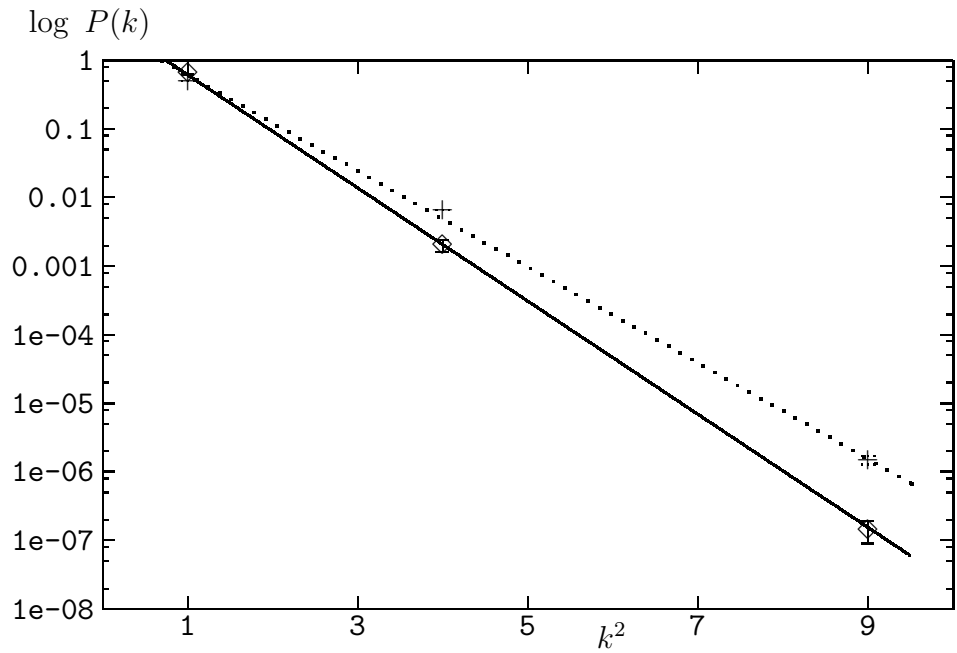
\begin{figure}
\centering
\setlength{\unitlength}{0.240900pt}
\ifx\plotpoint\undefined\newsavebox{\plotpoint}\fi
\sbox{\plotpoint}{\rule[-0.200pt]{0.400pt}{0.400pt}}%
\begin{picture}(1500,900)(0,0)
\font\gnuplot=cmtt10 at 12pt
\gnuplot
\sbox{\plotpoint}{\rule[-0.200pt]{0.400pt}{0.400pt}}%
\put(120.0,31.0){\rule[-0.200pt]{0.400pt}{207.656pt}}
\put(120.0,31.0){\rule[-0.200pt]{4.818pt}{0.400pt}}
\put(108,31){\makebox(0,0)[r]{1e-08}}
\put(1436.0,31.0){\rule[-0.200pt]{4.818pt}{0.400pt}}
\put(120.0,63.0){\rule[-0.200pt]{2.409pt}{0.400pt}}
\put(1446.0,63.0){\rule[-0.200pt]{2.409pt}{0.400pt}}
\put(120.0,106.0){\rule[-0.200pt]{2.409pt}{0.400pt}}
\put(1446.0,106.0){\rule[-0.200pt]{2.409pt}{0.400pt}}
\put(120.0,128.0){\rule[-0.200pt]{2.409pt}{0.400pt}}
\put(1446.0,128.0){\rule[-0.200pt]{2.409pt}{0.400pt}}
\put(120.0,139.0){\rule[-0.200pt]{4.818pt}{0.400pt}}
\put(108,139){\makebox(0,0)[r]{1e-07}}
\put(1436.0,139.0){\rule[-0.200pt]{4.818pt}{0.400pt}}
\put(120.0,171.0){\rule[-0.200pt]{2.409pt}{0.400pt}}
\put(1446.0,171.0){\rule[-0.200pt]{2.409pt}{0.400pt}}
\put(120.0,214.0){\rule[-0.200pt]{2.409pt}{0.400pt}}
\put(1446.0,214.0){\rule[-0.200pt]{2.409pt}{0.400pt}}
\put(120.0,236.0){\rule[-0.200pt]{2.409pt}{0.400pt}}
\put(1446.0,236.0){\rule[-0.200pt]{2.409pt}{0.400pt}}
\put(120.0,247.0){\rule[-0.200pt]{4.818pt}{0.400pt}}
\put(108,247){\makebox(0,0)[r]{1e-06}}
\put(1436.0,247.0){\rule[-0.200pt]{4.818pt}{0.400pt}}
\put(120.0,279.0){\rule[-0.200pt]{2.409pt}{0.400pt}}
\put(1446.0,279.0){\rule[-0.200pt]{2.409pt}{0.400pt}}
\put(120.0,322.0){\rule[-0.200pt]{2.409pt}{0.400pt}}
\put(1446.0,322.0){\rule[-0.200pt]{2.409pt}{0.400pt}}
\put(120.0,344.0){\rule[-0.200pt]{2.409pt}{0.400pt}}
\put(1446.0,344.0){\rule[-0.200pt]{2.409pt}{0.400pt}}
\put(120.0,354.0){\rule[-0.200pt]{4.818pt}{0.400pt}}
\put(108,354){\makebox(0,0)[r]{1e-05}}
\put(1436.0,354.0){\rule[-0.200pt]{4.818pt}{0.400pt}}
\put(120.0,387.0){\rule[-0.200pt]{2.409pt}{0.400pt}}
\put(1446.0,387.0){\rule[-0.200pt]{2.409pt}{0.400pt}}
\put(120.0,430.0){\rule[-0.200pt]{2.409pt}{0.400pt}}
\put(1446.0,430.0){\rule[-0.200pt]{2.409pt}{0.400pt}}
\put(120.0,452.0){\rule[-0.200pt]{2.409pt}{0.400pt}}
\put(1446.0,452.0){\rule[-0.200pt]{2.409pt}{0.400pt}}
\put(120.0,462.0){\rule[-0.200pt]{4.818pt}{0.400pt}}
\put(108,462){\makebox(0,0)[r]{1e-04}}
\put(1436.0,462.0){\rule[-0.200pt]{4.818pt}{0.400pt}}
\put(120.0,494.0){\rule[-0.200pt]{2.409pt}{0.400pt}}
\put(1446.0,494.0){\rule[-0.200pt]{2.409pt}{0.400pt}}
\put(120.0,537.0){\rule[-0.200pt]{2.409pt}{0.400pt}}
\put(1446.0,537.0){\rule[-0.200pt]{2.409pt}{0.400pt}}
\put(120.0,559.0){\rule[-0.200pt]{2.409pt}{0.400pt}}
\put(1446.0,559.0){\rule[-0.200pt]{2.409pt}{0.400pt}}
\put(120.0,570.0){\rule[-0.200pt]{4.818pt}{0.400pt}}
\put(108,570){\makebox(0,0)[r]{0.001}}
\put(1436.0,570.0){\rule[-0.200pt]{4.818pt}{0.400pt}}
\put(120.0,602.0){\rule[-0.200pt]{2.409pt}{0.400pt}}
\put(1446.0,602.0){\rule[-0.200pt]{2.409pt}{0.400pt}}
\put(120.0,645.0){\rule[-0.200pt]{2.409pt}{0.400pt}}
\put(1446.0,645.0){\rule[-0.200pt]{2.409pt}{0.400pt}}
\put(120.0,667.0){\rule[-0.200pt]{2.409pt}{0.400pt}}
\put(1446.0,667.0){\rule[-0.200pt]{2.409pt}{0.400pt}}
\put(120.0,678.0){\rule[-0.200pt]{4.818pt}{0.400pt}}
\put(108,678){\makebox(0,0)[r]{0.01}}
\put(1436.0,678.0){\rule[-0.200pt]{4.818pt}{0.400pt}}
\put(120.0,710.0){\rule[-0.200pt]{2.409pt}{0.400pt}}
\put(1446.0,710.0){\rule[-0.200pt]{2.409pt}{0.400pt}}
\put(120.0,753.0){\rule[-0.200pt]{2.409pt}{0.400pt}}
\put(1446.0,753.0){\rule[-0.200pt]{2.409pt}{0.400pt}}
\put(120.0,775.0){\rule[-0.200pt]{2.409pt}{0.400pt}}
\put(1446.0,775.0){\rule[-0.200pt]{2.409pt}{0.400pt}}
\put(120.0,785.0){\rule[-0.200pt]{4.818pt}{0.400pt}}
\put(108,785){\makebox(0,0)[r]{0.1}}
\put(1436.0,785.0){\rule[-0.200pt]{4.818pt}{0.400pt}}
\put(120.0,818.0){\rule[-0.200pt]{2.409pt}{0.400pt}}
\put(1446.0,818.0){\rule[-0.200pt]{2.409pt}{0.400pt}}
\put(120.0,861.0){\rule[-0.200pt]{2.409pt}{0.400pt}}
\put(1446.0,861.0){\rule[-0.200pt]{2.409pt}{0.400pt}}
\put(120.0,883.0){\rule[-0.200pt]{2.409pt}{0.400pt}}
\put(1446.0,883.0){\rule[-0.200pt]{2.409pt}{0.400pt}}
\put(120.0,893.0){\rule[-0.200pt]{4.818pt}{0.400pt}}
\put(108,893){\makebox(0,0)[r]{1}}
\put(1436.0,893.0){\rule[-0.200pt]{4.818pt}{0.400pt}}
\put(254.0,31.0){\rule[-0.200pt]{0.400pt}{4.818pt}}
\put(254,5){\makebox(0,0){1}}
\put(254.0,873.0){\rule[-0.200pt]{0.400pt}{4.818pt}}
\put(521.0,31.0){\rule[-0.200pt]{0.400pt}{4.818pt}}
\put(521,5){\makebox(0,0){3}}
\put(521.0,873.0){\rule[-0.200pt]{0.400pt}{4.818pt}}
\put(788.0,31.0){\rule[-0.200pt]{0.400pt}{4.818pt}}
\put(788,5){\makebox(0,0){5}}
\put(788.0,873.0){\rule[-0.200pt]{0.400pt}{4.818pt}}
\put(1055.0,31.0){\rule[-0.200pt]{0.400pt}{4.818pt}}
\put(1055,5){\makebox(0,0){7}}
\put(1055.0,873.0){\rule[-0.200pt]{0.400pt}{4.818pt}}
\put(1322.0,31.0){\rule[-0.200pt]{0.400pt}{4.818pt}}
\put(1322,5){\makebox(0,0){9}}
\put(1322.0,873.0){\rule[-0.200pt]{0.400pt}{4.818pt}}
\put(120.0,31.0){\rule[-0.200pt]{321.842pt}{0.400pt}}
\put(1456.0,31.0){\rule[-0.200pt]{0.400pt}{207.656pt}}
\put(120.0,893.0){\rule[-0.200pt]{321.842pt}{0.400pt}}
{\large
\put(100,950){\makebox(0,0){$\log\ P(k)$}}
\put(920,-10){\makebox(0,0){$k^2$}}
}
\put(120.0,31.0){\rule[-0.200pt]{0.400pt}{207.656pt}}
\put(254,872){\raisebox{-.8pt}{\makebox(0,0){$\Diamond$}}}
\put(654,602){\raisebox{-.8pt}{\makebox(0,0){$\Diamond$}}}
\put(1322,154){\raisebox{-.8pt}{\makebox(0,0){$\Diamond$}}}
\put(254.0,871.0){\usebox{\plotpoint}}
\put(244.0,871.0){\rule[-0.200pt]{4.818pt}{0.400pt}}
\put(244.0,872.0){\rule[-0.200pt]{4.818pt}{0.400pt}}
\put(654.0,592.0){\rule[-0.200pt]{0.400pt}{4.577pt}}
\put(644.0,592.0){\rule[-0.200pt]{4.818pt}{0.400pt}}
\put(644.0,611.0){\rule[-0.200pt]{4.818pt}{0.400pt}}
\put(1322.0,134.0){\rule[-0.200pt]{0.400pt}{8.431pt}}
\put(1312.0,134.0){\rule[-0.200pt]{4.818pt}{0.400pt}}
\put(1312.0,169.0){\rule[-0.200pt]{4.818pt}{0.400pt}}
\put(254,861){\makebox(0,0){$+$}}
\put(654,658){\makebox(0,0){$+$}}
\put(1322,265){\makebox(0,0){$+$}}
\put(254,861){\usebox{\plotpoint}}
\put(254,861){\usebox{\plotpoint}}
\put(244.00,861.00){\usebox{\plotpoint}}
\put(264,861){\usebox{\plotpoint}}
\put(244.00,861.00){\usebox{\plotpoint}}
\put(264,861){\usebox{\plotpoint}}
\put(654,658){\usebox{\plotpoint}}
\put(654,658){\usebox{\plotpoint}}
\put(644.00,658.00){\usebox{\plotpoint}}
\put(664,658){\usebox{\plotpoint}}
\put(644.00,658.00){\usebox{\plotpoint}}
\put(664,658){\usebox{\plotpoint}}
\put(1322.00,258.00){\usebox{\plotpoint}}
\put(1322,271){\usebox{\plotpoint}}
\put(1312.00,258.00){\usebox{\plotpoint}}
\put(1332,258){\usebox{\plotpoint}}
\put(1312.00,271.00){\usebox{\plotpoint}}
\put(1332,271){\usebox{\plotpoint}}
\sbox{\plotpoint}{\rule[-0.400pt]{0.800pt}{0.800pt}}%
\multiput(220.00,891.09)(0.751,-0.500){1549}{\rule{1.402pt}{0.120pt}}
\multiput(220.00,891.34)(1166.090,-778.000){2}{\rule{0.701pt}{0.800pt}}
\sbox{\plotpoint}{\rule[-0.500pt]{1.000pt}{1.000pt}}%
\multiput(212,893)(18.071,-10.210){66}{\usebox{\plotpoint}}
\put(1389,228){\usebox{\plotpoint}}
\end{picture}
\vskip 5mm
\caption{Estimated probabilities for more than $k$
Incipient Spanning Clusters as function of $k^2$ in logarithmic
scale. The dotted line is a fit for lattices with free boundaries
and the solid line is a fit for the lattices with periodic boundaries
in vertical direction.}
\label{pk2}
\end{figure}

\newpage

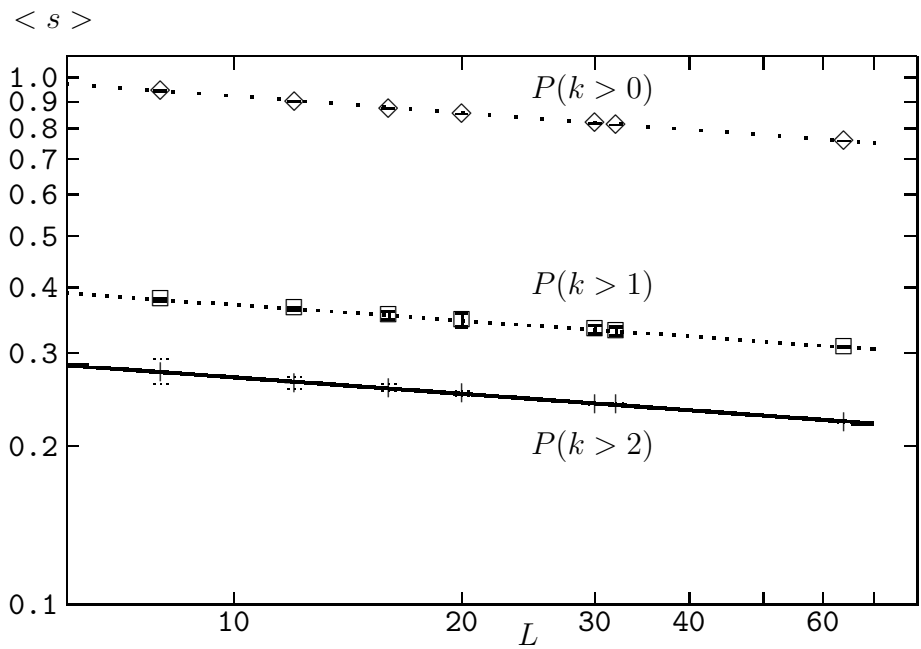
\begin{figure}
\centering
\setlength{\unitlength}{0.240900pt}
\ifx\plotpoint\undefined\newsavebox{\plotpoint}\fi
\sbox{\plotpoint}{\rule[-0.200pt]{0.400pt}{0.400pt}}%
\begin{picture}(1500,900)(0,0)
\font\gnuplot=cmtt10 at 12pt
\gnuplot
\sbox{\plotpoint}{\rule[-0.200pt]{0.400pt}{0.400pt}}%
\put(120.0,31.0){\rule[-0.200pt]{4.818pt}{0.400pt}}
\put(108,31){\makebox(0,0)[r]{0.1}}
\put(1436.0,31.0){\rule[-0.200pt]{4.818pt}{0.400pt}}
\put(120.0,280.0){\rule[-0.200pt]{4.818pt}{0.400pt}}
\put(108,280){\makebox(0,0)[r]{0.2}}
\put(1436.0,280.0){\rule[-0.200pt]{4.818pt}{0.400pt}}
\put(120.0,426.0){\rule[-0.200pt]{4.818pt}{0.400pt}}
\put(108,426){\makebox(0,0)[r]{0.3}}
\put(1436.0,426.0){\rule[-0.200pt]{4.818pt}{0.400pt}}
\put(120.0,529.0){\rule[-0.200pt]{4.818pt}{0.400pt}}
\put(108,529){\makebox(0,0)[r]{0.4}}
\put(1436.0,529.0){\rule[-0.200pt]{4.818pt}{0.400pt}}
\put(120.0,610.0){\rule[-0.200pt]{4.818pt}{0.400pt}}
\put(108,610){\makebox(0,0)[r]{0.5}}
\put(1436.0,610.0){\rule[-0.200pt]{4.818pt}{0.400pt}}
\put(120.0,675.0){\rule[-0.200pt]{4.818pt}{0.400pt}}
\put(108,675){\makebox(0,0)[r]{0.6}}
\put(1436.0,675.0){\rule[-0.200pt]{4.818pt}{0.400pt}}
\put(120.0,731.0){\rule[-0.200pt]{4.818pt}{0.400pt}}
\put(108,731){\makebox(0,0)[r]{0.7}}
\put(1436.0,731.0){\rule[-0.200pt]{4.818pt}{0.400pt}}
\put(120.0,779.0){\rule[-0.200pt]{4.818pt}{0.400pt}}
\put(108,779){\makebox(0,0)[r]{0.8}}
\put(1436.0,779.0){\rule[-0.200pt]{4.818pt}{0.400pt}}
\put(120.0,821.0){\rule[-0.200pt]{4.818pt}{0.400pt}}
\put(108,821){\makebox(0,0)[r]{0.9}}
\put(1436.0,821.0){\rule[-0.200pt]{4.818pt}{0.400pt}}
\put(120.0,859.0){\rule[-0.200pt]{4.818pt}{0.400pt}}
\put(108,859){\makebox(0,0)[r]{1.0}}
\put(1436.0,859.0){\rule[-0.200pt]{4.818pt}{0.400pt}}
\put(120.0,893.0){\rule[-0.200pt]{4.818pt}{0.400pt}}
\put(1436.0,893.0){\rule[-0.200pt]{4.818pt}{0.400pt}}
\put(383.0,31.0){\rule[-0.200pt]{0.400pt}{4.818pt}}
\put(383,5){\makebox(0,0){10}}
\put(383.0,873.0){\rule[-0.200pt]{0.400pt}{4.818pt}}
\put(741.0,31.0){\rule[-0.200pt]{0.400pt}{4.818pt}}
\put(741,5){\makebox(0,0){20}}
\put(741.0,873.0){\rule[-0.200pt]{0.400pt}{4.818pt}}
\put(950.0,31.0){\rule[-0.200pt]{0.400pt}{4.818pt}}
\put(950,5){\makebox(0,0){30}}
\put(950.0,873.0){\rule[-0.200pt]{0.400pt}{4.818pt}}
\put(1098.0,31.0){\rule[-0.200pt]{0.400pt}{4.818pt}}
\put(1098,5){\makebox(0,0){40}}
\put(1098.0,873.0){\rule[-0.200pt]{0.400pt}{4.818pt}}
\put(1214.0,31.0){\rule[-0.200pt]{0.400pt}{4.818pt}}
\put(1214.0,873.0){\rule[-0.200pt]{0.400pt}{4.818pt}}
\put(1308.0,31.0){\rule[-0.200pt]{0.400pt}{4.818pt}}
\put(1308,5){\makebox(0,0){60}}
\put(1308.0,873.0){\rule[-0.200pt]{0.400pt}{4.818pt}}
\put(1387.0,31.0){\rule[-0.200pt]{0.400pt}{4.818pt}}
\put(1387.0,873.0){\rule[-0.200pt]{0.400pt}{4.818pt}}
\put(1456.0,31.0){\rule[-0.200pt]{0.400pt}{4.818pt}}
\put(1456.0,873.0){\rule[-0.200pt]{0.400pt}{4.818pt}}
\put(120.0,31.0){\rule[-0.200pt]{321.842pt}{0.400pt}}
\put(1456.0,31.0){\rule[-0.200pt]{0.400pt}{207.656pt}}
\put(120.0,893.0){\rule[-0.200pt]{321.842pt}{0.400pt}}
{\large
\put(100,950){\makebox(0,0){$<s>$}}
\put(845,-15){\makebox(0,0){$L$}}
\put(850,840){\makebox(0,0)[l]{$P(k>0)$}}
\put(850,532){\makebox(0,0)[l]{$P(k>1)$}}
\put(850,280){\makebox(0,0)[l]{$P(k>2)$}}
}
\put(120.0,31.0){\rule[-0.200pt]{0.400pt}{207.656pt}}
\put(268,838){\raisebox{-.8pt}{\makebox(0,0){$\Diamond$}}}
\put(478,821){\raisebox{-.8pt}{\makebox(0,0){$\Diamond$}}}
\put(626,810){\raisebox{-.8pt}{\makebox(0,0){$\Diamond$}}}
\put(741,802){\raisebox{-.8pt}{\makebox(0,0){$\Diamond$}}}
\put(950,787){\raisebox{-.8pt}{\makebox(0,0){$\Diamond$}}}
\put(983,784){\raisebox{-.8pt}{\makebox(0,0){$\Diamond$}}}
\put(1341,759){\raisebox{-.8pt}{\makebox(0,0){$\Diamond$}}}
\put(268.0,837.0){\rule[-0.200pt]{0.400pt}{0.482pt}}
\put(258.0,837.0){\rule[-0.200pt]{4.818pt}{0.400pt}}
\put(258.0,839.0){\rule[-0.200pt]{4.818pt}{0.400pt}}
\put(478.0,821.0){\usebox{\plotpoint}}
\put(468.0,821.0){\rule[-0.200pt]{4.818pt}{0.400pt}}
\put(468.0,822.0){\rule[-0.200pt]{4.818pt}{0.400pt}}
\put(626.0,810.0){\usebox{\plotpoint}}
\put(616.0,810.0){\rule[-0.200pt]{4.818pt}{0.400pt}}
\put(616.0,811.0){\rule[-0.200pt]{4.818pt}{0.400pt}}
\put(741,802){\usebox{\plotpoint}}
\put(731.0,802.0){\rule[-0.200pt]{4.818pt}{0.400pt}}
\put(731.0,802.0){\rule[-0.200pt]{4.818pt}{0.400pt}}
\put(950.0,786.0){\usebox{\plotpoint}}
\put(940.0,786.0){\rule[-0.200pt]{4.818pt}{0.400pt}}
\put(940.0,787.0){\rule[-0.200pt]{4.818pt}{0.400pt}}
\put(983,784){\usebox{\plotpoint}}
\put(973.0,784.0){\rule[-0.200pt]{4.818pt}{0.400pt}}
\put(973.0,784.0){\rule[-0.200pt]{4.818pt}{0.400pt}}
\put(1341,759){\usebox{\plotpoint}}
\put(1331.0,759.0){\rule[-0.200pt]{4.818pt}{0.400pt}}
\put(1331.0,759.0){\rule[-0.200pt]{4.818pt}{0.400pt}}
\put(268,397){\makebox(0,0){$+$}}
\put(478,379){\makebox(0,0){$+$}}
\put(626,371){\makebox(0,0){$+$}}
\put(741,363){\makebox(0,0){$+$}}
\put(950,346){\makebox(0,0){$+$}}
\put(983,346){\makebox(0,0){$+$}}
\put(1341,318){\makebox(0,0){$+$}}
\multiput(268,377)(0.000,20.756){2}{\usebox{\plotpoint}}
\put(268,417){\usebox{\plotpoint}}
\put(258.00,377.00){\usebox{\plotpoint}}
\put(278,377){\usebox{\plotpoint}}
\put(258.00,417.00){\usebox{\plotpoint}}
\put(278,417){\usebox{\plotpoint}}
\put(478.00,369.00){\usebox{\plotpoint}}
\put(478,388){\usebox{\plotpoint}}
\put(468.00,369.00){\usebox{\plotpoint}}
\put(488,369){\usebox{\plotpoint}}
\put(468.00,388.00){\usebox{\plotpoint}}
\put(488,388){\usebox{\plotpoint}}
\put(626.00,366.00){\usebox{\plotpoint}}
\put(626,377){\usebox{\plotpoint}}
\put(616.00,366.00){\usebox{\plotpoint}}
\put(636,366){\usebox{\plotpoint}}
\put(616.00,377.00){\usebox{\plotpoint}}
\put(636,377){\usebox{\plotpoint}}
\put(741.00,359.00){\usebox{\plotpoint}}
\put(741,366){\usebox{\plotpoint}}
\put(731.00,359.00){\usebox{\plotpoint}}
\put(751,359){\usebox{\plotpoint}}
\put(731.00,366.00){\usebox{\plotpoint}}
\put(751,366){\usebox{\plotpoint}}
\put(950.00,344.00){\usebox{\plotpoint}}
\put(950,348){\usebox{\plotpoint}}
\put(940.00,344.00){\usebox{\plotpoint}}
\put(960,344){\usebox{\plotpoint}}
\put(940.00,348.00){\usebox{\plotpoint}}
\put(960,348){\usebox{\plotpoint}}
\put(983.00,344.00){\usebox{\plotpoint}}
\put(983,347){\usebox{\plotpoint}}
\put(973.00,344.00){\usebox{\plotpoint}}
\put(993,344){\usebox{\plotpoint}}
\put(973.00,347.00){\usebox{\plotpoint}}
\put(993,347){\usebox{\plotpoint}}
\put(1341.00,317.00){\usebox{\plotpoint}}
\put(1341,318){\usebox{\plotpoint}}
\put(1331.00,317.00){\usebox{\plotpoint}}
\put(1351,317){\usebox{\plotpoint}}
\put(1331.00,318.00){\usebox{\plotpoint}}
\put(1351,318){\usebox{\plotpoint}}
\sbox{\plotpoint}{\rule[-0.400pt]{0.800pt}{0.800pt}}%
\put(268,510){\raisebox{-.8pt}{\makebox(0,0){$\Box$}}}
\put(478,496){\raisebox{-.8pt}{\makebox(0,0){$\Box$}}}
\put(626,485){\raisebox{-.8pt}{\makebox(0,0){$\Box$}}}
\put(741,477){\raisebox{-.8pt}{\makebox(0,0){$\Box$}}}
\put(950,463){\raisebox{-.8pt}{\makebox(0,0){$\Box$}}}
\put(983,460){\raisebox{-.8pt}{\makebox(0,0){$\Box$}}}
\put(1341,435){\raisebox{-.8pt}{\makebox(0,0){$\Box$}}}
\put(268.0,508.0){\rule[-0.400pt]{0.800pt}{0.964pt}}
\put(258.0,508.0){\rule[-0.400pt]{4.818pt}{0.800pt}}
\put(258.0,512.0){\rule[-0.400pt]{4.818pt}{0.800pt}}
\put(478.0,494.0){\rule[-0.400pt]{0.800pt}{0.964pt}}
\put(468.0,494.0){\rule[-0.400pt]{4.818pt}{0.800pt}}
\put(468.0,498.0){\rule[-0.400pt]{4.818pt}{0.800pt}}
\put(626.0,479.0){\rule[-0.400pt]{0.800pt}{2.891pt}}
\put(616.0,479.0){\rule[-0.400pt]{4.818pt}{0.800pt}}
\put(616.0,491.0){\rule[-0.400pt]{4.818pt}{0.800pt}}
\put(741.0,467.0){\rule[-0.400pt]{0.800pt}{5.059pt}}
\put(731.0,467.0){\rule[-0.400pt]{4.818pt}{0.800pt}}
\put(731.0,488.0){\rule[-0.400pt]{4.818pt}{0.800pt}}
\put(950.0,456.0){\rule[-0.400pt]{0.800pt}{3.132pt}}
\put(940.0,456.0){\rule[-0.400pt]{4.818pt}{0.800pt}}
\put(940.0,469.0){\rule[-0.400pt]{4.818pt}{0.800pt}}
\put(983.0,454.0){\rule[-0.400pt]{0.800pt}{3.132pt}}
\put(973.0,454.0){\rule[-0.400pt]{4.818pt}{0.800pt}}
\put(973.0,467.0){\rule[-0.400pt]{4.818pt}{0.800pt}}
\put(1341.0,435.0){\usebox{\plotpoint}}
\put(1331.0,435.0){\rule[-0.400pt]{4.818pt}{0.800pt}}
\put(1331.0,436.0){\rule[-0.400pt]{4.818pt}{0.800pt}}
\sbox{\plotpoint}{\rule[-0.500pt]{1.000pt}{1.000pt}}%
\multiput(120,520)(20.706,-1.438){62}{\usebox{\plotpoint}}
\put(1387,432){\usebox{\plotpoint}}
\sbox{\plotpoint}{\rule[-0.600pt]{1.200pt}{1.200pt}}%
\multiput(120.00,404.26)(6.920,-0.500){174}{\rule{16.826pt}{0.120pt}}
\multiput(120.00,404.51)(1232.077,-92.000){2}{\rule{8.413pt}{1.200pt}}
\sbox{\plotpoint}{\rule[-0.500pt]{1.000pt}{1.000pt}}%
\put(120,849){\usebox{\plotpoint}}
\multiput(120,849)(41.400,-3.039){31}{\usebox{\plotpoint}}
\put(1387,756){\usebox{\plotpoint}}
\end{picture}
\vskip 5mm
\caption{Mean size of $k$ simultaneous Incipient Spanning Clusters, measured on
the corresponding events. The slope of the curves is close to the exact
value of $\beta/\nu$ for the all plotted curves. 
}
\label{size}
\end{figure}

\end{document}